# Analysis of some solutions to protect the western tombolo of Giens

Van Van THAN[1], Yves LACROIX[2,3], Pierre LIARDET

**Corresponding author:** Y. Lacroix, yves.lacroix@univ-tln.fr

**Dedicatory:** *in memory of Pierre Liardet, who left us September 2014, whom we miss and appreciated.*

**Abstract.** The tombolo of Giens is located in the town of Hyères (France). We recall the history of coastal erosion, and proeminent factors affecting the evolution of the western tombolo. We then discuss the possibility of stabilizing the western tombolo. Our argumentation relies on a coupled model integrating swells, currents, water levels and sediment transport. We present the conclusions of the simulations of various scenarios, including pre-existing propositions from coastal engineering offices. We conclude that beach replenishment seems to be necessary but not sufficient for the stabilization of the beach. Breakwaters reveal effective particularly in the most exposed northern area. Some solutions fulfill conditions so as to be elected as satisfactory. We give a comparative analysis of the efficiency of 14 alternatives for the protection of the tombolo.

**Keywords**: tombolo, replenishment, silting, breakwaters, coastal erosion, evolution, coupled models.

1. **Introduction.**

The geographic coordinates of the tombolo of Giens are 43.039615°N to 43.081654°N and 6.125244°E to 6.156763°E, between the Gulf of Giens and Hyères harbor. The Alamanarre beach consists of the western part of the tombolo , and is subject to coastal erosion. Since more than 50 years research has been conducted to try to understand the dynamics of this erosion (Blanc 1973, Grissac 1975), and help establishing a protection plan for the coast which presents important economical and enviromental impacts. In a preceding paper we collected all available data on the subject, compoiled it to numeric format, and calibrated a coupled model using MIKE 21 so as to understand the proeminent factors at the origin of this erosion process, and the hydro-sedimentologal dynamics of this complex system (Than, Lacroix et al. 2014).

There it was concluded that the tombolo should be divided into four significant cells and that heavy impact occured mainly during southwestern winter storm events conjugated to atmospheric depression.

In the present note we use our calibrated model and investigate numerically some solution proposals, some arising from engineering consultants, others we propose here. We also give a look to economical aspects of the proposed solutions (costs), our aim being to help politics make a decision towards this recurrent and yet costy problem (beach replenishment occurs each year and costs more than $3 \times 10^5$ € a year, without maintaining stable the situation which worthens).

The cells are determined by their limiting landmarks: North (A) is from north boundary to B03, North-central (B) B03 to B16, central (C) B16 to B23, and South (D) B23 to B46 (Serantoni and Lizaud 2000-2010). The North-central zone is the most affected by erosion (Grissac 1975, CETE 1992, Courtaud 2000). The absence of natural sediment supply and ancient anthropic influence weigh also on the erosion process (Blanc 1974, HYDRO-M 1993). The coastline has driven back east by 50m to 80m in the central zone, 75m to 90m in the southern, since 1956 (Blanc 1974); by 15m to 20m in the north zone (average 40 years, 1950 to 1998) (Courtaud 2000). The height of the sand dune has decreased by 0.3 to 1.5m in the late 80's and the 90's (Blanc 1975).

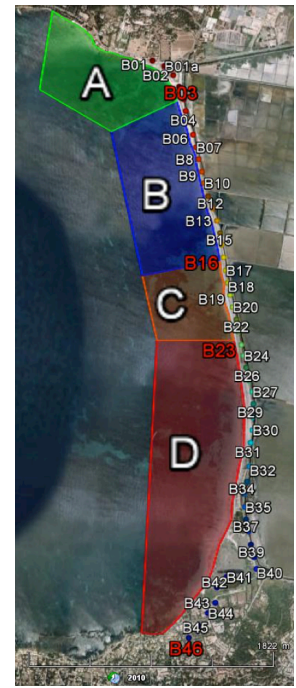

*Figure 1: western tombolo cells*

[1] AMUE, laboratoire LATP
[2] UTLN, SEATECH, avenue G. Pompidou, 83162 La Valette du Var, France
[3] MEMOCS, University of l'Aquila, Italy

*Table 1: cells for the western tombolo of Giens*

| Cell | Mean width (m) | Length (m) | Perimeter (m) | Surface (m$^2$) |
|---|---|---|---|---|
| A | 533 | 1.257 | 3.074 | 517.070 |
| B | 585 | 1.287 | 3.578 | 703.840 |
| C | 570 | 626 | 2.316 | 325.655 |
| D | 623 | 2.544 | 5.332 | 1.273.597 |
| Total | 2.311 | 5.714 | 14.300 | 1.721.343 |

Hereafter we will describe previous attempts to protect the tombolo. Then, we investigate soft solutions, solutions with underwater structures, and combined solutions, using our calibrated model. We will estimate there efficiency by the change of global bathymetric volume on the zone of study, and the evolution of the coast profile at different landmarks. We will also investigate the efficiency towards the transport of sediments between cells. And finally we shall estimate the costs for each solution.

We have tested 15 scenarios, among which number 1 corresponds to status quo. The estimation is based on different regimes we identified as characteristic in the previous article, for the data period of year 2008.

The paper ends with a conclusive section.

## 2. Protection of the tombolo: the proposed alternatives.
### 2.1. Previous attempts

The northern beach has been locally protected by a riprap revetment that was recently removed (Courtaud 2000). Indeed, the effect of these blocks on the beach sedimentological balance was negative. After a storm in 1994, which destroyed the dune and submerged the parallel road ("route du sel"), the choice was made to start periodically replenishing the dune (Courtaud 2000). Later, ganivelles were installed (so as to protect from anthropic destruction, the area being highly touristic summer time), and car parkings were organized (Courtaud 2000). Also, the "route du sel" was closed winter time.

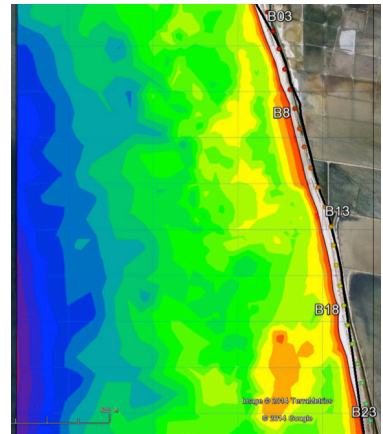

*Figure 2: bathymetry alternative 0*

The area is protected (Conservatoire du Littoral) and close to a high environmental importance zone (Parc Naturel de Port Cros). This means that any solution to be proposed should take into account visual, environmental, and economical impacts.

Relaoding occurs winter time essentially but reveals non sufficient and costy. Each year the dune is restored (with a mixture of sand and posidonia leaves) but this does not stop the beach drawback in the north central zone.

Alternative 0 will be for us status quo: just go on as it is.

### 2.2. Soft solutions

The submerged area is covered mainly by a posidonia field, which absorbs wave energy, retains offshore sediment transport, and covers during western and south western episodes the beach with posidonia algae that has the property to damp wave impacts to the coast. Recently the decision was taken to preserve this algae coverage, though touristic attractivity has little decreased. The preservation of the posidonia field and the maintain of ganivelles has already limited the erosion process, though the limitation has not been precisely measured.

Replenishment with a sand-algae mixture allows to maintain and restaure the dune in the north-central and north zones.

### 2.3. Pure silting scenarios

Silting with quarry sand or gravels has been proposed by (ERAMM 2001). It can be envisaged with or without structures. ERAMM proposed reinforcing beach foot by the way, arguing that without additional structure the solution is short term. Here is the description of alternatives 1 and 2, consisting of only silting with quarry sand in restricted areas. The bathymetry comes from our previous paper.

*Table 2: silting volumes for alternatives 1 and 2*

| Alternative | Protection | V (m³) | Length | Width |
|---|---|---|---|---|
| 1 | B07 to B11 | 66680 | 460m | 136m |
| 2 | North+North-central | 218061 | 2000m | 136m |

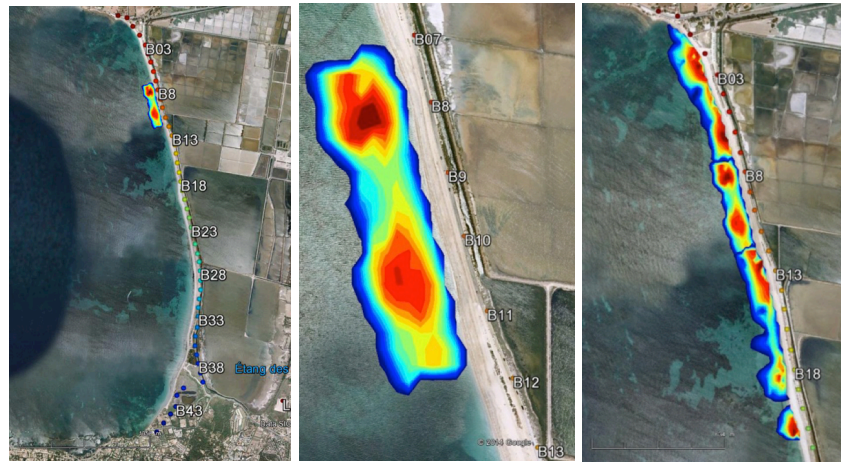

*Figure 3: bathymetry for alternative 1 (left, centre) and 2 (right).*

## 2.4. Adding small structures to 1 and 2

The next alternatives (3 to 6) will consist in silting and adding a beach foot and immersed breakwaters, or immersed breakwaters with immersed groins (ERAMM 2001).

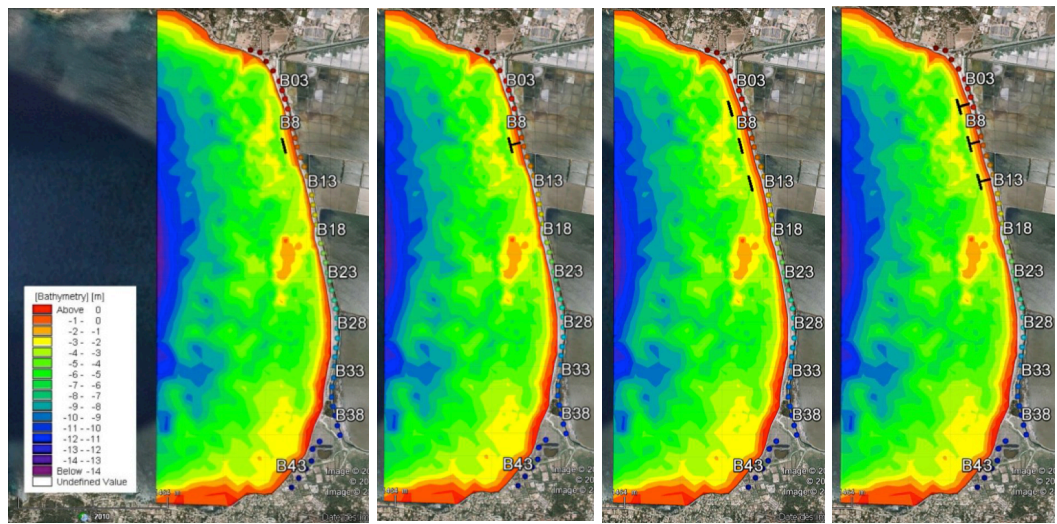

*Figure 4: left to right alternatives 3 to 6, silting with small immersed structures*

*Table 3: silting and small structures – characteristics for alternatives 3-6*

| Alternative | | 3 | 4 | 5 | 6 |
|---|---|---|---|---|---|
| Protection | | B07 to B11 | B07 to B11 | North-central | North-central |
| Silting volume | V (m3) | 66680 | 66680 | 218061 | 218061 |
| | Length | 460 | 460 | 2000 | 2000 |
| | Width | 136 | 136 | 136 | 136 |
| Beach foot | Quantity | 1 | 1 | 3 | 3 |
| | Crest freeboard (m) | -1,12 | -1,12 | -1,12 | -1,12 |
| | Shore dist. | 100 | 100 | 100 | 100 |

|  | Alternative | 3 | 4 | 5 | 6 |
|---|---|---|---|---|---|
|  | Length | 150 | 150 | 150 | 150 |
|  | Crest width | 230 | 230 | 230 | 230 |
|  | Spacing | 12 | 12 | 12 | 12 |
| Immersed breakwater and groin | Quantity | None | 1 | None | 3 |
|  | Crest freeboard (m) | - | -1,12 to 0 | - | De -1,12 to 0 |
|  | Length | - | 100 | - | 100 |
|  | Crest width | - | 380 | - | 380 |
|  | Spacing | - | 12 | - | 12 |

## 2.5. Hard coastal protection solutions

We propose protection solutions with two bareers of breakwaters, one close and one further away offshore (Blanc 1973, SOGREAH 1988). Breakwaters will be either made of concrete, either from riprap (SOGREAH 1988). In the model the two alternatives are treated as solid. The following table gives a description of scenarios 7 to 11:

*Table 4: characteristics for structures alternatives 7-11*

|  | Alternative | 7 | 8 | 9 | 10 | 11 |
|---|---|---|---|---|---|---|
|  | Protection | B07 to B11 | North-central | North-central | Whole beach | Whole beach |
| Offshore immersed breakwaters | Quantity | None | 2 | 2 | 4 | 4 |
|  | Crest freeboard | - | -3 | -3 | -3 | -3 |
|  | Shore dist. | - | 400 | 400 | 400 | 400 |
|  | Length | - | 340 | 340 | 340 | 340 |
|  | Spacing | - | 280 | 280 | 280 | 280 |
|  | Crest width | - | 12 | 12 | 12 | 12 |
| Close shore immersed breakwaters | Quantity | 1 | 3 | 3 | 6 | 6 |
|  | Crest freeboard | -2 | -2 | -2 | -2 | -2 |
|  | Shore dist. | 200 | 300 | 300 | 300 | 300 |
|  | Length | 440 | 340 | 340 | 340 | 340 |
|  | Spacing | - | 280 | 280 | 280 | 280 |
|  | Crest width | 12 | 12 | 3 | 3 | 3 |
| Immersed breakwater with groin | Quantity | 2 | None | 3 | None | 3 |
|  | Crest freeboard | De -2 à 0 | - | De -2 à 0 | - | De -2 à 0 |
|  | Length | 200 | - | 300 | - | 300 |
|  | Spacing | 440 | - | 620 | - | 620 |
|  | Crest width | 12 | - | 12 | - | 12 |

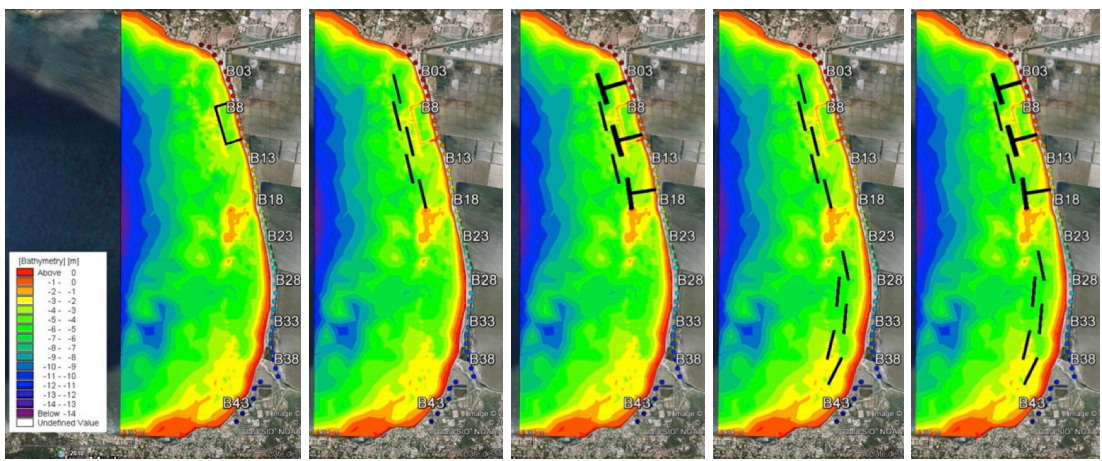

*Figure 5: alternatives 7 to 11*

## 2.6. Combined alternatives (soft and hard)

The combined alternatives (12, 13, 14) are are as follows: 12 is as 7 but with beach replenishment from landmark B07 to B11. And 13 and 14 are respectively as 8 and 9, with the same replenishing as 12.

## 3. Scenarios evaluation via numerical simulation

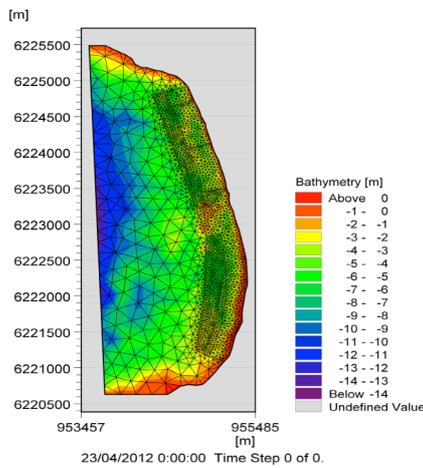

*Figure 6: domain and grid structure.*

*Table 5: characteristics of alternatives 0-14: summary*

| Alt. | Silting (m³) | Structure length (m) | | | |
|---|---|---|---|---|---|
| | | Beach foot | Immersed breakwater +groin | Immersed breakwater | |
| | | | | Close | Far |
| 0 | Dune preservation | | | | |
| 1 | 66680 | | | | |
| 2 | 218061 | | | | |
| 3 | 66680 | 150 | | | |
| 4 | 66680 | 150 | 100 | | |
| 5 | 218061 | 450 | | | |
| 6 | 218061 | 450 | 300 | | |
| 7 | | | 400 | 440 | |
| 8 | | | | 1020 | 680 |
| 9 | | | 900 | 1020 | 680 |
| 10 | | | | 2040 | 1360 |
| 11 | | | 900 | 2040 | 1360 |
| 12 | 66680 | | 400 | 440 | |
| 13 | 66680 | | | 1020 | 680 |
| 14 | 66680 | | 900 | 1020 | 680 |

Our model couples wave, current, and sediment transport (Tang, Keen et al. 2009). It was calibrated as described in our previous paper. Immersed structures are simply represented by a change in bathymetry, with mean grain size equal to $D_{50}$=20cm which makes it « solid » (unaffected by sediment transport, Manning and rugosity are those of rock).

### 3.1. Domain

The grid is made of triangles with 3635 triangles and 2000 nodes, with increased model resolution at the proximity of immersed structures when are present. See Figure 6 above.

### 3.2. The evaluation of the efficiency of a scenario

The model is run over two significant types of conditions as defined in our previous paper. One is average yearly based on 2008 observations, which are relevant for the 1995-2010 period. The other is of type tempest with south-western conditions which are the worse for erosion phenomena, but also western conditions and mistral episodes. These conditions are used as border conditions for the grid.

#### 3.2.1. Global volume change of the area

The results are collected in the following table: the changes are evaluated per cell.

*Table 6: efficiency evaluated by average m³/day change per zone*

| Zone \ Alt. | Annual conditions: average m³/day change per zone | | | | | Tempest conditions; average m³/day change per zone | | | | |
|---|---|---|---|---|---|---|---|---|---|---|
| | North | N-central | Central | South | Total | North | N-central | Central | South | Total |
| 0 | -52 | -663 | 4 | -13 | -724 | -1162 | -10141 | -1477 | -3297 | -16077 |
| 1 | -103,8 | 117,6 | -88 | 125,9 | 51,7 | -851 | -5271 | -2157 | -3814 | -12093 |
| 2 | -105,9 | 74,4 | -88,6 | 115,2 | -4,9 | -874 | -5273 | -2106 | -3781 | -12034 |
| 3 | -106,4 | 131 | -69,4 | 134,8 | 90 | -960 | -5341 | -2150 | -3879 | -12330 |
| 4 | -103,9 | 136,3 | -59,6 | 137,9 | 111 | -946 | -5293 | -2144 | -3877 | -12260 |
| 5 | -95,5 | -18,7 | 18,3 | 124,8 | 28,9 | -984 | -5645 | -2020 | -3874 | -12523 |
| 6 | -93,6 | -26 | 6,7 | 126,4 | 13,5 | -803 | -5623 | -2033 | -3847 | -12306 |
| 7 | 167,9 | 171,9 | -104,3 | 711,8 | 947 | -784 | -5717 | -1886 | -4174 | -12561 |
| 8 | 14,6 | 149,2 | -1,1 | -34,4 | 128 | -874 | -4409 | -1746 | -5779 | -12808 |
| 9 | -50,5 | 151 | -18,9 | 63,3 | 145 | -859 | -4407 | -1607 | -5972 | -12845 |
| 10 | -49,5 | 185 | -26,9 | 86,3 | 195 | -972 | -4387 | -1817 | -4999 | -12175 |
| 11 | -51,4 | 156,5 | -20,6 | 81,3 | 166 | -757 | -4403 | -1643 | -5106 | -11909 |
| 12 | 173,1 | 163,2 | -107,1 | 705,7 | 935 | -796 | -5706 | -1961 | -4272 | -12735 |
| 13 | -3,7 | 140 | -42,9 | 67,1 | 161 | -880 | -4455 | -1804 | -5718 | -12857 |
| 14 | -26,8 | 127,4 | -50 | 70,2 | 121 | -819 | -4388 | -1693 | -5925 | -12825 |

Here are graphic versions of the above table:

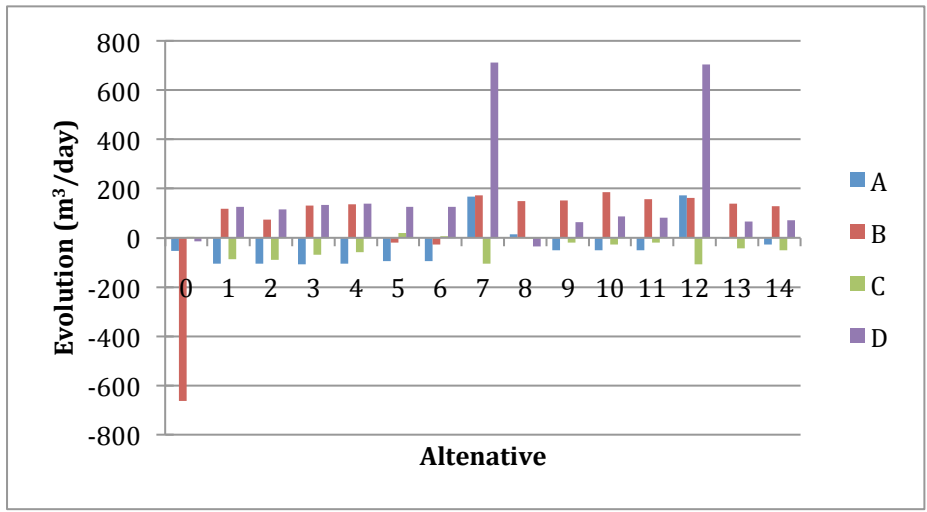

*Figure 7: volume variation per cell annual conditions*

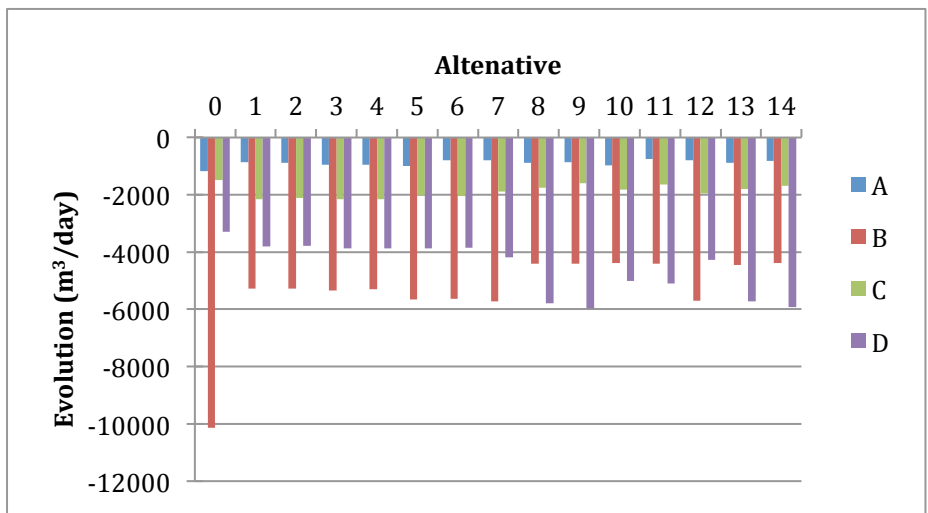

*Figure 8: volume variation per cell storm conditions*

3.2.2. <u>Beach profil evolution</u>.

We have extracted from the simulation the beach profile evolution at landmark B08 with reference point alternative 0. First we look at the alternative's effect under annual conditions, and we group alternatives by type: the first group is alternatives 1 to 6 with reference point alternative 0; the second group is alternatives 7 to 14 with reference point alternative 0:

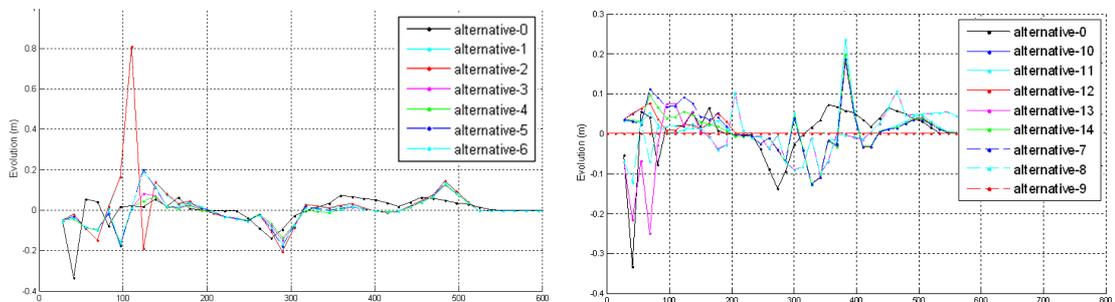

*Figure 9: change in bathymetry landmark B08 annual conditions*

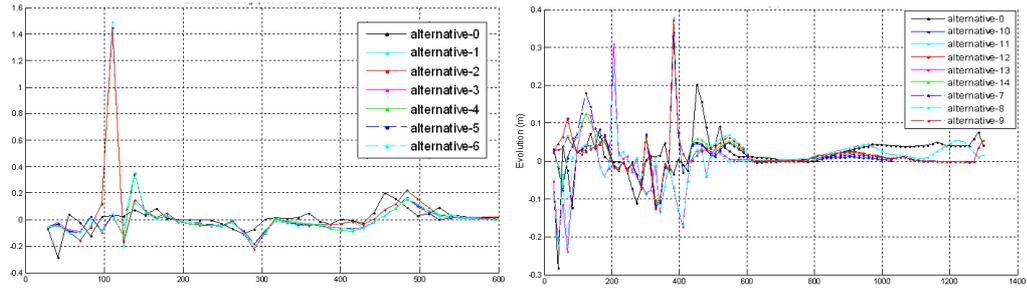

*Figure 10: change in bathymetry landmark B08 storm conditions*

### 3.2.3. Sediment transport per cell

*Table 7: sediment transport per zone unit is $10^{-6} m^3/s/lm$*

| Zone Alternative | Annual conditions | | | | Storm conditions | | | |
|---|---|---|---|---|---|---|---|---|
| | A | B | C | D | A | B | C | D |
| 0  | 0,5 | 3,9 | 0,9 | 1,4 | 4,6 | 20,3 | 15,4 | 16,4 |
| 9  | 0,3 | 2,1 | 2,4 | 1,6 | 2,2 | 12,4 | 12,5 | 12,8 |
| 10 | 0,2 | 2,1 | 2,2 | 1,5 | 2,6 | 13,8 | 13,6 | 11,2 |
| 11 | 0,3 | 2,2 | 2,4 | 1,5 | 2,3 | 12,6 | 12,4 | 11,4 |
| 12 | 0,4 | 3,1 | 2,2 | 5,4 | 1,4 | 14,1 | 18,8 | 16,3 |
| 13 | 0,3 | 2,4 | 2,3 | 1,6 | 2,0 | 13,1 | 13,7 | 12,6 |
| 14 | 0,3 | 2,5 | 2,5 | 1,6 | 2,0 | 13,2 | 12,7 | 12,8 |
| 1  | 0,3 | 2,2 | 2,4 | 1,7 | 1,5 | 11,1 | 23,4 | 12,7 |
| 2  | 0,3 | 2,6 | 2,5 | 1,7 | 1,5 | 11,7 | 23,3 | 12,9 |
| 3  | 0,3 | 2,1 | 2,4 | 1,7 | 1,7 | 10,5 | 23,8 | 12,8 |
| 4  | 0,3 | 2,1 | 2,4 | 1,7 | 1,7 | 10,7 | 23,9 | 12,7 |
| 5  | 0,3 | 2,3 | 2,7 | 1,7 | 1,9 | 10,2 | 24,4 | 12,9 |
| 6  | 0,3 | 2,6 | 2,7 | 1,7 | 1,6 | 10,6 | 24,4 | 12,9 |
| 7  | 0,4 | 2,7 | 2,2 | 5,4 | 1,3 | 13,3 | 18,1 | 16,3 |
| 8  | 0,5 | 0,6 | 1,2 | 1,4 | 2,4 | 13,4 | 13,8 | 12,6 |

For better readability, Table 6 is resumed by the two following graphics:

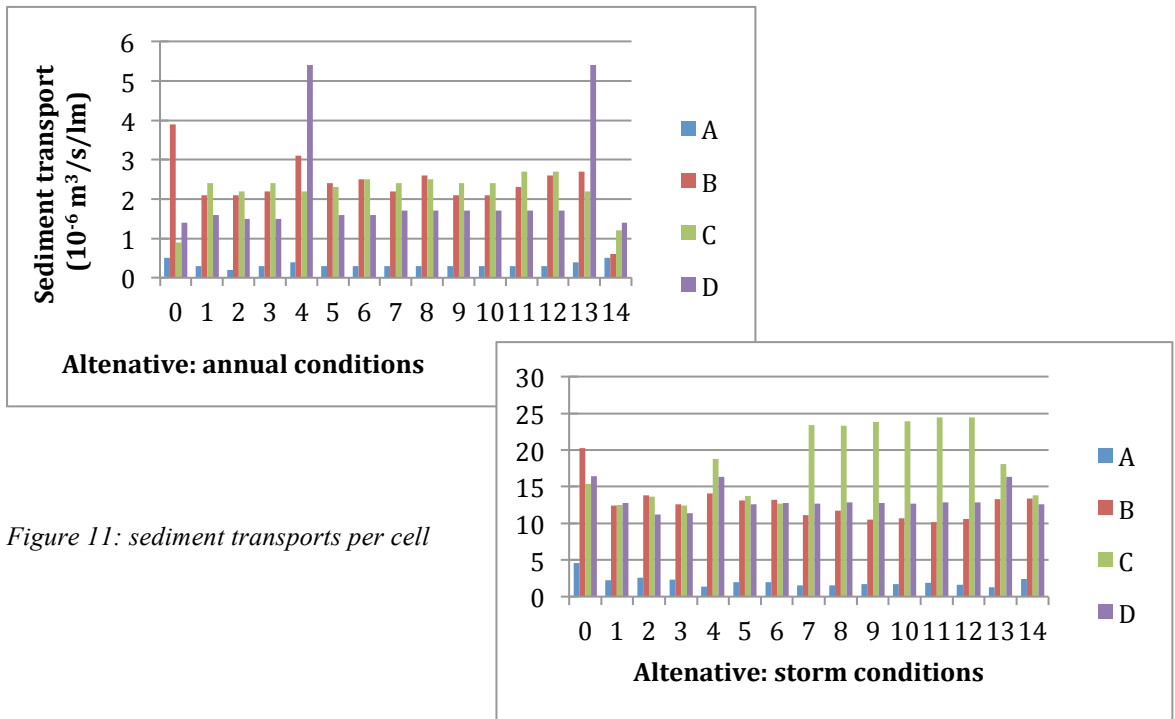

*Figure 11: sediment transports per cell*

Sediment transport has also been extracted on profile B08 for the different alternatives. We observe (see figures below) that sediment transport decreases between 100m and 400m. The alternatives 10, 11, 13 and 14 even generate accretion at this sensible profile.

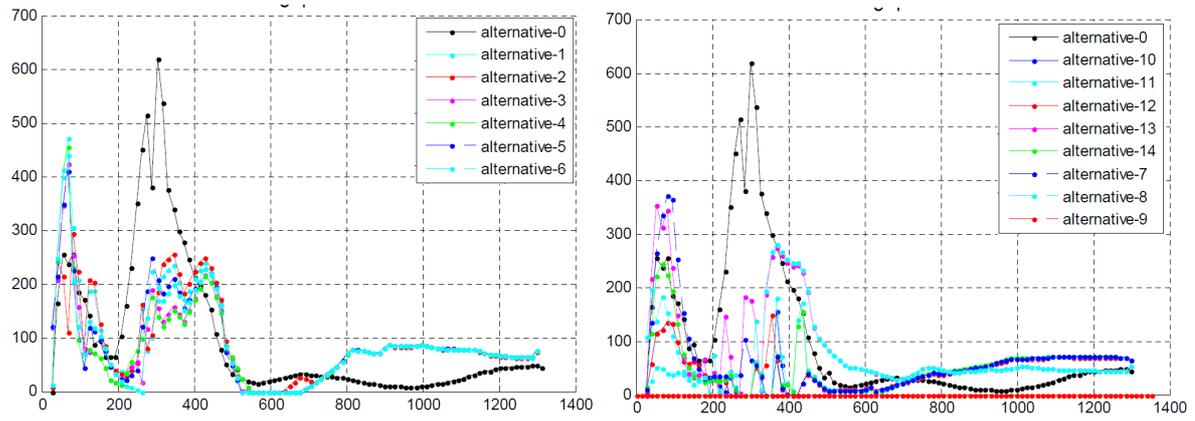

*Figure 12: sediment transport m3/year/lm profile B08 annual conditions*

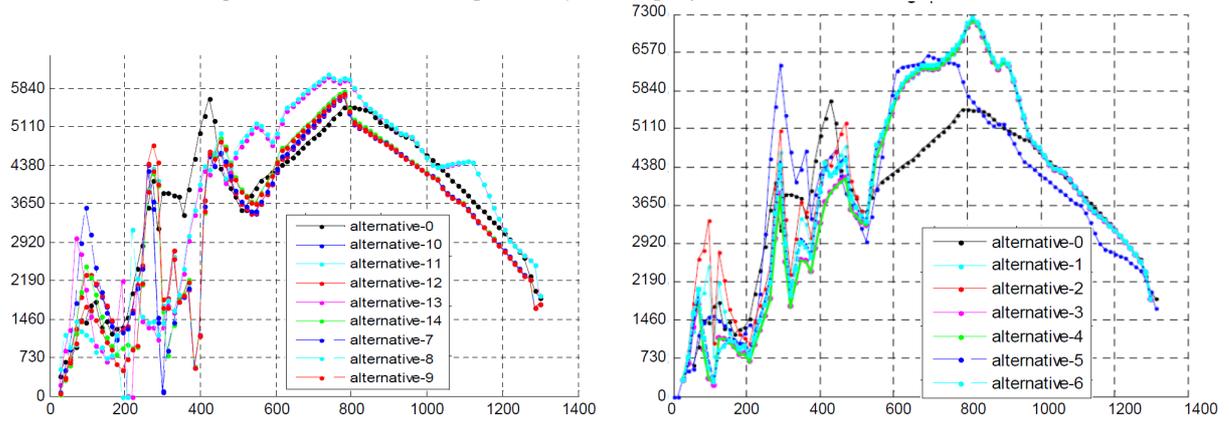

*Figure 13: sediment transport m3/year/lm profile B08 storm conditions*

### 3.2.4. Wave attenuation
#### 3.2.4.1. Wave height et period attenuation

We consider $K_{t1} = H_t / H_i$ for wave heights attenuation coefficient (CERC 1984, Liao, Jiang et al. 2013), and $K_{t2} = T_t / T_i$ for period attenuation, where subscripts «i» reffer to initial (offshore conditions), and «t» reffers to terminal (at the coast). An extraction point close to the coast (-1m) and one offshore (-3m) have been defined, at which the model gave the following results:

*Table 8: wave attenuation coefficients*

| Alt. | Annual | | | | | | Storm | | | | | |
|---|---|---|---|---|---|---|---|---|---|---|---|---|
| | Height (m) | | | Period (s) | | | Height(m) | | | Period (s) | | |
| | $H_i$ | $H_t$ | $K_{t1}$ | $T_i$ | $T_t$ | $K_{t2}$ | $H_i$ | $H_t$ | $K_{t1}$ | $T_i$ | $T_t$ | $K_{t2}$ |
| 0 | 0,8 | 0,7 | 0,88 | 4,5 | 4 | 0,89 | 1,6 | 1,1 | 0,69 | 7,7 | 7,1 | 0,92 |
| 1 | 0,8 | 0,7 | 0,88 | 4,5 | 3,9 | 0,87 | 1,6 | 1 | 0,63 | 7,7 | 7 | 0,91 |
| 2 | 0,8 | 0,7 | 0,88 | 4,5 | 3,9 | 0,87 | 1,6 | 1 | 0,63 | 7,7 | 7 | 0,91 |
| 3 | 0,8 | 0,6 | 0,75 | 4,5 | 3,8 | 0,84 | 1,6 | 0,9 | 0,56 | 7,7 | 6,9 | 0,90 |
| 4 | 0,8 | 0,6 | 0,75 | 4,5 | 3,8 | 0,84 | 1,6 | 0,9 | 0,56 | 7,7 | 6,9 | 0,90 |
| 5 | 0,8 | 0,6 | 0,75 | 4,5 | 3,8 | 0,84 | 1,6 | 0,9 | 0,56 | 7,7 | 6,9 | 0,90 |
| 6 | 0,8 | 0,6 | 0,75 | 4,5 | 3,8 | 0,84 | 1,6 | 0,9 | 0,56 | 7,7 | 6,9 | 0,90 |
| 7 | 0,8 | 0,7 | 0,88 | 4,4 | 3,9 | 0,89 | 1,6 | 1 | 0,63 | 7,7 | 7,1 | 0,92 |
| 8 | 0,8 | 0,6 | 0,75 | 4,5 | 3,9 | 0,87 | 1,6 | 1 | 0,63 | 7,7 | 7,1 | 0,92 |
| 9 | 0,8 | 0,6 | 0,75 | 4,5 | 3,9 | 0,87 | 1,6 | 1 | 0,63 | 7,7 | 7,1 | 0,92 |
| 10 | 0,8 | 0,6 | 0,75 | 4,5 | 3,9 | 0,87 | 1,6 | 1 | 0,63 | 7,7 | 7,1 | 0,92 |
| 11 | 0,8 | 0,6 | 0,75 | 4,5 | 3,9 | 0,87 | 1,6 | 1 | 0,63 | 7,7 | 7,1 | 0,92 |
| 12 | 0,8 | 0,6 | 0,75 | 4,4 | 3,8 | 0,86 | 1,6 | 0,9 | 0,56 | 7,7 | 7 | 0,91 |
| 13 | 0,8 | 0,6 | 0,75 | 4,5 | 3,8 | 0,84 | 1,6 | 0,9 | 0,56 | 7,7 | 7 | 0,91 |
| 14 | 0,8 | 0,6 | 0,75 | 4,5 | 3,8 | 0,84 | 1,6 | 0,9 | 0,56 | 7,7 | 7 | 0,91 |

We observe here that alternatives 3, 4, 5, 6 and 12, 13 and 14 are the most efficient for wave attenuation.

#### 3.2.4.2. Energy density attenuation

We have extracted from the simulation the density of energy distribution for the four cells with reference point alternative 0.

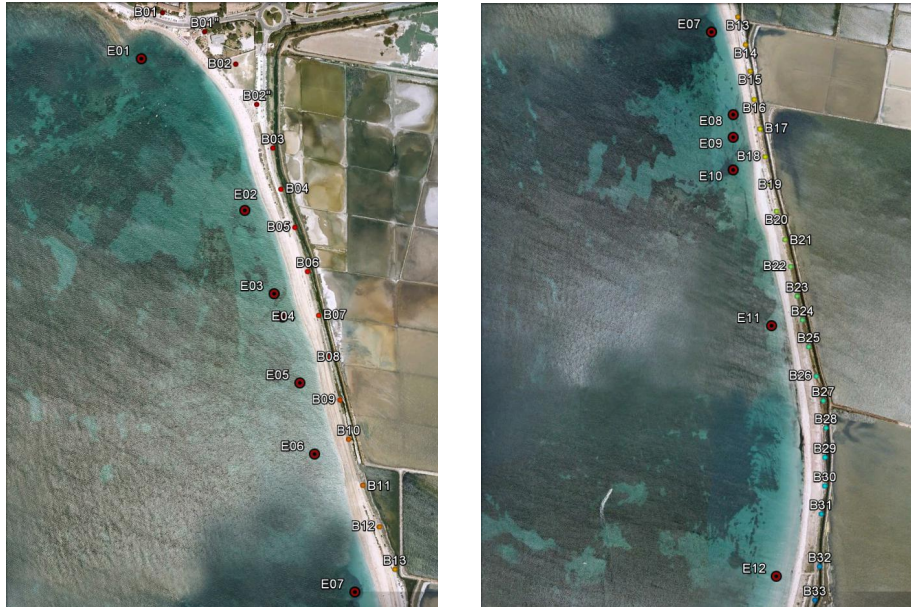

*Figure 14:* extraction point close to the coast

An extraction point close to the coast (-1m) from E01 to E12 has been defined (fig. 14), at which the model gave the following results:

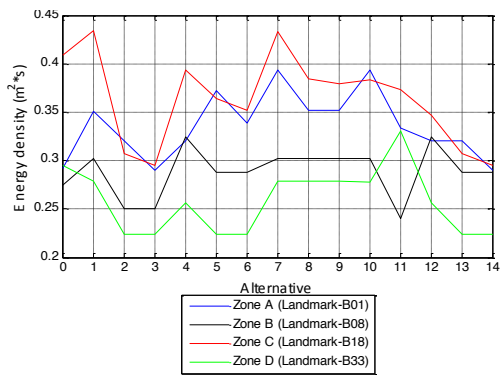 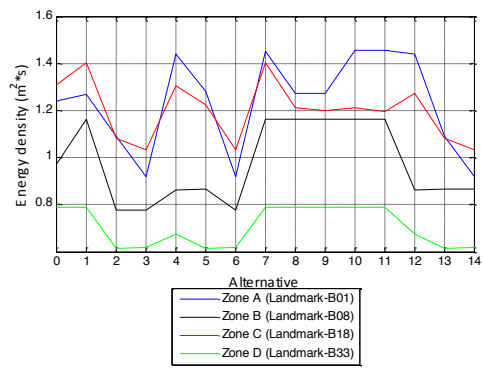

*Figure 15: density energy distribution annual conditions*

*Figure 16: density energy distribution storm conditions*

We also have extracted the density of energy distribution for the North zone (see fig. 16, 17).

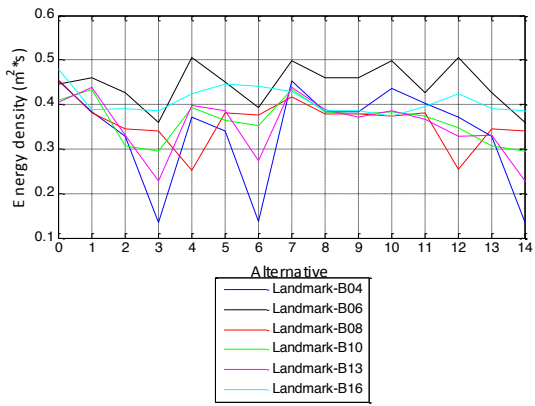 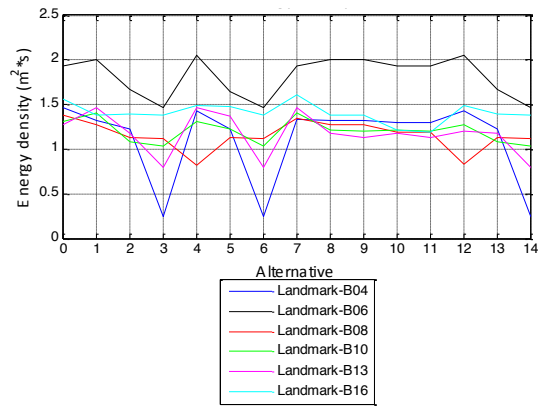

*Figure 17: distribution energy density in North zone annual conditions*

*Figure 18: distribution energy density in North zone storm conditions*

We observe here that alternatives 2, 3, 5, 6, and 12, 13 and 14 are the most efficient for energy density attenuation.

### 3.2.5. Cost effectiveness

The estimation of cost is based on estimation of costs of materials, and structures by linear meter of length. Other costs should be taken into account as follow up costs for bathymetry, communication. However these extra costs will not be integrated in our calculus.

This estimation is based upon the following materials and construction prices:

*Table 9: costs for the alternatives (Alt.) 1-14*

| Alt. | Silting (m$^3$) | Length (m) | | | | Cost (€ before tax) | | |
|---|---|---|---|---|---|---|---|---|
| | | Beach foot | Immersed breakwater + groin | Immersed breakwater | | Construction | Maintenance | Total |
| | | | | Close shore | Second row | | | |
| 1 | 66680 | | | | | 733 480 | 88 000 | 821 480 |
| 3 | 66680 | 150 | | | | 1 114 630 | 88 000 | 1 202 630 |
| 4 | 66680 | 150 | 100 | | | 1 368 730 | 88 000 | 1 456 730 |
| 7 | | | | 400 | 440 | 2 134 440 | | 2 134 440 |
| 2 | 218061 | | | | | 2 398 671 | 88 000 | 2 486 671 |
| 12 | 66680 | | | 400 | 440 | 2 867 920 | 88 000 | 2 955 920 |
| 5 | 218061 | 450 | | | | 3 542 121 | 88 000 | 3 630 121 |
| 8 | | | | 1020 | 680 | 4 319 700 | | 4 319 700 |
| 6 | 218061 | 450 | 300 | | | 4 304 421 | 88 000 | 4 392 421 |
| 13 | 66680 | | | 1020 | 680 | 5 053 180 | - | 5 053 180 |
| 9 | | | 900 | 1020 | 680 | 6 606 600 | | 6 606 600 |
| 14 | 66680 | | 900 | 1020 | 680 | 7 340 080 | - | 7 340 080 |
| 10 | | | | 2040 | 1360 | 8 639 400 | | 8 639 400 |
| 11 | | | 900 | 2040 | 1360 | 10 926 300 | | 10 926 300 |

*Table 10: materials costs per unit and type of operation*

| ID | Technic | Equipment / materials | Unit | Cost (€ before tax) | | Source |
|---|---|---|---|---|---|---|
| | | | | Installation | Maintenance | |
| **1** | **Average materials cost** | | | | | |
| 1.1 | Sand and gravel | On shore | T | 15,24 | | ERAMM, 2001 |
| 1.2 | Sand and gravel | Pumping and discharge | T | 7,62 | | |
| 1.3 | ballast | By sea | T | 21,34 | | |
| 1.4 | rockfill | | m3 | 76,22 | | |
| | | | lm | 4065 | | |
| 1.5 | pile | | lm | 2541 | | |
| 1.6 | geotextile | | | 2541 | | |
| 1.7 | sand | | T | 12,2 | | SOGREAH, 1988 |
| | | | m3 | 24,39 | | |
| **2** | **Cost of solutions** | | | | | |
| 2.1 | Replenishment | sand | m3 | 7 à 14 | Highly variable | |
| | | granular | lm | 45 | | DDE[1] 13 |
| 2.2 | Beach reprofiling | | lm | 70 | | EID Med |
| 2.3 | Replenishment | On shore | lm | 4574 à 5336 | | ERAMM, 2001 |
| | | Maritim transport | lm | 6098 | | |
| 2.4 | Dune creation | | lm | 320 à 400 | 20 € /lm ganivelles | EID Med[2] |
| 2.5 | Dune maintenance and restauration | | lm | 75 | | EID Med |
| 2.6 | Breakwater + groin | | lm | 2500 | Each year: 3 à 5% of initial price | DDE 13 |
| 2.7 | Immersed breakwater | | lm | 4000 | | BCEOM[3] |
| 2.8 | Semi-immersed breakwater | | lm | 6200 | | |
| 2.9 | Beach foot | geotextile | lm | 2000 | Each year: 3 à 5% of initial price | BCEOM, 2004 |
| | | rockfill | lm | 4000 | | |
| | | geotextile | lm | 700 | | ERAMM, 2004 |
| | | rockfill | lm | 1330 | | |
| | | geotextile | lm | 1100 | | BRL[4], 2005 |
| | | geotextile | lm | 7200 | | DDE13, 2006 |
| | | rockfill | lm | 1143 | | ERAMM, 2001 |
| | | pile | lm | 686 | | |
| | | geotextile | lm | 686 | | |

---

[1] DDE 13: Direction Départementale de l'Equipement des Bouches du Rhône
[2] EID Med: Entente Interdépartementale de Démoustication Méditerranée
[3] Bureau Central d'Etudes pour les équipements d'Outre-Mer
[4] compagnie d'aménagement du Bas-Rhône et du Languedoc

The protection of Western tombolo in Giens requires the construction of structures. Otherwise the « route du sel » will disappear and this may cause practical problems for circulation and security regarding the access to the presqu'île de Giens. As predicted, without intervention, the coast line should drawback from 15 to 40m.

The city of Hyères was thinking about implementing a new road, tracing in the middle, north to south, of the tombolo. An analysis of 35 projects of road constructions in France between 1997 and 2002 reveals a mean cost for a standard road of 5 M€ BT/km (Cazala, Deterne et al. 2006). Thus the cost of a new road would be of at least 20M€ BT, without taking into account environmental impacts (natural protected zone) and visual impacts.

Compared to, our solutions imply costs ranging from 0.9M€ BT to 11 M€ BT.

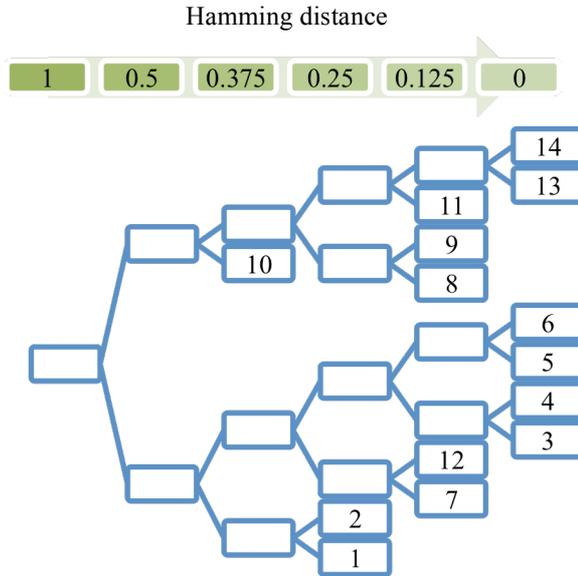

*Figure 19: non oriented cluster analysis of alternatives versus criteria using Hamming distance*

*Table 11: criteria*

| 1 | Volume change |
|---|---|
| 2 | Profile change and sediment transport |
| 3 | Wave reduction |
| 4 | Dune preservation |
| 5 | Soft solution |
| 6 | Cost |
| 7 | Environmental impact |
| 8 | Impact on recreational activities |

*Table 12: alternative's marks for criteria*

| Alt. | Criteria | | | | | | | |
|---|---|---|---|---|---|---|---|---|
|  | 1 | 2 | 3 | 4 | 5 | 6 | 7 | 8 |
| 1 | + | - | - | - | + | + | + | + |
| 2 | - | - | - | - | + | + | + | + |
| 3 | + | - | ++ | - | + | - | - | + |
| 4 | + | - | ++ | - | + | - | - | + |
| 5 | - | - | ++ | - | + | - | - | + |
| 6 | - | - | ++ | - | + | - | - | --- |
| 7 | + | - | - | - | - | - | - | -- |
| 8 | + | - | + | + | - | -- | -- | - |
| 9 | + | - | + | + | - | -- | -- | -- |
| 10 | + | + | + | + | - | -- | -- | - |
| 11 | + | + | + | + | - | -- | --- | -- |
| 12 | + | - | ++ | - | - | - | - | - |
| 13 | + | + | ++ | + | - | -- | -- | -- |
| 14 | + | + | ++ | + | - | -- | -- | -- |

4. **Recommendations for the protection.**

We first observe that the results of simulation, whatever the alternative, conform to the general knowledge of the study area. The choice of an optimal solution among the alternatives should be driven by taking into account several criteria (Table 10). We then give a mark towards criteria for each alternative, from « ++ » (very good) to « --- » (very very bad) (SOGREAH 1988). The preceding section helps us completing the following table:

We observe that silting and replenishment seem unavoidable for the preservation of the beach.

We use a matlab statistic toolbox to create clusters from the Table 11 matrix, using hamming distance in the pdist function and then applying linkage.

We could of course ponderate the criteria and drive different clusters, but at this point we are making no decision so this is reserved for future work.

Figure 19 pictures the clusters from Table 11, there are 3 clusters at 0 Hamming distance (13-14, 3-4, 5-6), 5 clusters at distance 0.125, one of which groups solutions 11, 13, 14. Going further, at distance 0.25 we have 6 clusters, and at distance 0.375 (meaning any two cluster members differ at 3/8 coordinates in Table 11) we obtain 4 clusters (8-9-11-13-14, 10, 3-4-5-6-7-12, 1-2). Finally, we end with two clusters on the 50% disagreement level, 8-9-10-11-13-14, and 1-2-3-4-5-6-7-12.

If we wish to push on the erosion criteria alternatives 10, 11, 13, 14 reveal comparable and best performance in this aspect.

## 5. References.

Blanc, J. J. (1973). Recherches sédimentologiques sur la protection du littoral à la presqu'ile de Giens (Var). Centre d'Océanologie d'Endoume, CNEXO**:** 47.

Blanc, J. J. (1974). "Phénomènes d'érosions sous-marines de la presqu'île de Giens (Var)." C.R.Acad. Sci. Paris **278**: 1821-1823.

Blanc, J. J. (1975). Recherche de sédimentologie appliquée au littoral rocheux de la Provence. Aménagement et protection, CNEXO**:** 164.

Cazala, A., J. Deterne, G. Crespy, P. Garnier, G. D. Monchy and P. Rimattei (2006). Rapport sur la comparaison au niveau européen des coûts de construction, d'entretien et d'exploitation des routes, Contrôle général économique et financier & Conseil général des ponts.

CERC (1984). Shore Protection Manual. U. G. P. Office. US Army Corps of Engineers, Coastal Engineering Research Center.

CETE (1992). Etudes de dunes. Protection du tombolo ouest de Giens. DDE du Var. protection du tombolo ouest de Giens, DDE Var.

Courtaud, J. (2000). Dynamiques géomorphologiques et risques littoraux cas du tombolo de giens (Var, France méridionale) Thèse de Doctorat, Université de Provence.

ERAMM (2001). Etude sur la protection de la partie Nord du tombolo Ouest de Giens, City of Hyères. **phase I+II+III**.

Grissac, J. D. (1975). Sédimentologie dynamique des rades d'Hyères et de Giens (Var)- Problèmes d'aménagements Thèse de doctorat, Université de Provence.

HYDRO-M (1993). Etude d'impact sur l'environnement du projet de protection du tombolo ouest de la presqu'ile de Giens, City of Hyères.

Liao, Y. C., J. H. Jiang, Y. P. Wu and C. P. Lee (2013). "Experimental study of wave breaking criteria and energy loss caused by a submerged porous breakwater on horizontal bottom." Journal of Marine Science and Technology **21**(1): 35-41.

Serantoni, P. and O. Lizaud (2000-2010). Suivi de l'évolution des plages de la commune Hyères-les-palmiers, City of Hyères.

SOGREAH (1988). Protection du tombolo Ouest, City of Hyères. **Rapport 2-1a**.

SOGREAH (1988). Protection du tombolo Ouest.

Tang, H. S., T. R. Keen and R. Khanbilvardi (2009). "A model-coupling framework for nearshore waves, currents, sediment transport, and seabed morphology." Commun Nonlinear Sci Numer Simulat **14**: 2935–2947.

Than, Lacroix, Liardet and Léandri (2014). Analysis of a coupled hydro-sedimentological numerical model for the western tombolo of Giens. SEATECH/LATP/MEMOCS/IGS**:** 10.